\documentclass{IOS-Book-Article}

\usepackage{mathptmx}

\usepackage{url}%

\begin{document}
\begin{frontmatter}    
\title{Hybrid Intelligence for \\Digital Humanities}

 \author{\fnms{Victor} \snm{de Boer}} \and
 \runningauthor{V. de Boer and L. Stork}\author{\fnms{Lise} \snm{Stork}}
 \address{{Vrije Universiteit Amsterdam, the Netherlands. \{v.de.boer, l.stork\}@vu.nl}}

\begin{abstract}
In this paper, we explore the synergies between Digital Humanities (DH) as a discipline and Hybrid Intelligence (HI) as a research paradigm. In DH research, the use of digital methods and specifically that of Artificial Intelligence is subject to a set of requirements and constraints. We argue that these are well-supported by the capabilities and goals of HI. Our contribution includes the identification of five such DH requirements: Successful AI systems need to be able to 1) collaborate with the (human) scholar; 2) support data criticism; 3) support tool criticism; 4) be aware of and cater to various perspectives and 5) support distant and close reading. We take the CARE principles of Hybrid Intelligence (collaborative, adaptive, responsible and explainable) as theoretical framework and map these to the DH requirements. In this mapping, we include example research projects. 
We finally address how insights from DH can be applied to HI and discuss open challenges 
for the combination of the two disciplines. 
\end{abstract}

\end{frontmatter}

\thispagestyle{empty}
\pagestyle{empty}

\section{Introduction}

Digital tools and methods have permeated all fields of science, including the humanities. This has given rise to the area of scholarly activity known as Digital Humanities (DH)\cite{haigh2014we}. With more museum, archive and library collections made available in digital form, either through digitization or as `digital born', questions arise on how digital tools of exploration, analysis, annotation and visualisation can be incorporated in the humanities workflows and methodologies. Here, it has been recognized that the use of such tools can make this research more effective and efficient and open up new possibilities for analysing data or telling stories. At the same time, these tools and interdisciplinary methods will need to be carefully matched to the values of the humanities research process with respect to transparency and reproducible methods \cite{berry2012introduction}.

More recently, Artificial Intelligence (AI) tools and methods have become more available also to stakeholders in the field of humanities. Researchers from both AI and humanities have been identifying potential challenges and opportunities for AI in Digital Humanities. These opportunities include the possibility to analyse heritage and other humanities data at scale \cite{Gefen2021}, open up new ways of modelling humanities knowledge, analysing data and recognizing patterns, but also disseminating results \cite{pavlidis2022ai}. Challenges identified in literature include issues around bias in data and/or AI algorithms \cite{berry2022ai, heritage7020038} as well as challenges and ethics of categorization~\cite{jansen2019ontologies}, or the lack of polyvocality in representation~\cite{van2021polyvocal}. Concerns have also been expressed around the issue of diversity, equity and inclusion (DEI) in the larger field of AI itself~\cite{chun2023crisis}. Such investigations call for more human-centric approaches to AI to be developed that match Digital Humanities requirements. We argue here that the recently introduced research paradigm of Hybrid Intelligence \cite{akata2020research} aligns well with these requirements.  

In this paper, we explore the synergies between DH as a discipline and HI as a research paradigm. We identify that in Digital Humanities research, the use of digital methods and specifically that of AI is subject to a set of requirements and constraints that are well-supported by the capabilities and goals of Hybrid Intelligence. We take the CARE principles of Hybrid Intelligence (collaborative, adaptive, responsible and explainable) as theoretical framework and map these to the DH needs of tool and data criticism, transparency and ethical considerations. We also address how insights from DH can be applied to challenges identified in the context of Hybrid Intelligence, specifically where it concerns developing responsible and transparent systems. Through a description of example DH projects, we further illustrate this synergy. We finally identify open challenges and opportunities for the combination of the two disciplines.

\section{Challenges for the use of AI in Digital Humanities}
\label{sec:dh}
    \subsection{Understanding Digital Humanities}
Digital humanities (DH) represents a scholarly realm where computing or digital technologies intersect with the various disciplines within the humanities. This field encompasses the methodical utilization of digital resources in humanities research, coupled with the examination of their practical applications\cite{drucker2021digital}. DH is characterized by innovative approaches to scholarly endeavors, involving collaborative, transdisciplinary, and computationally oriented practices in research, teaching, and publishing\cite{burdick2016digital_humanities}. By introducing digital tools and methodologies into humanities studies, DH acknowledges the shift away from the printed word as the primary medium for knowledge production and dissemination.
DH researchers employ a diverse set of methodologies and tools to analyze, interpret, and present data. Textual analysis stands as a cornerstone, leveraging computational techniques such as natural language processing (NLP) and sentiment analysis to dissect literary works, historical documents, and cultural texts. Additionally, DH embraces network analysis, utilizing graph theory to map relationships among entities, uncovering hidden connections within datasets, and shedding light on complex societal structures. Visualization tools play a pivotal role, translating intricate data into visual representations that aid in comprehending patterns, and relationships within information. 

\subsection{AI in DH}
Within Digital Humanities, there has been quite a lot of interest in the application of Artificial Intelligence technologies for annotation, analysis and dissemination of DH research. Various types of AI solutions are being applied to specific DH challenges.
\textit{Knowledge Representation} approaches have been used to capture and model data, information and knowledge from various domains. Especially Semantic Web principles and practices have been embraced by the galleries, archive, libraries and museum (GLAM) community, to connect heterogeneous and distributed collections\cite{hyvonen2009culturesampo,deBoer2013amsterdam}. By using standard data models, interoperability is increased and standard ways of reasoning can be applied. At the same time, representation of heritage data as knowledge graphs also allows for other downstream analyses such as through machine learning or visualisation. 
\textit{Machine Learning} has been used to analyse large humanities datasets, 
with a recent focus on exploring deep learning approaches. Challenges for the use of AI in DH also include the limited extent to which current Machine Learning solutions can be used for multimodal or sparse datasets (e.g. \cite{lallensack2022machine,stork2021large}). 
Critical views on the role of Machine Learning in DH have been part of scholarly debate, focusing on bias, transparency and the way that usage of ML could influence the methodological rigour of humanities research \cite{bassett2017critical}.
The humanities focuses on artifacts produced by human agents, and the majority of these artifacts have historically been natural language texts. \textit{Natural Language Processing} (NLP) therefore takes an important role in DH. Most NLP pipelines now contain at least some Machine Learning, either using pre-trained (large) language models (LLMs) or using models specified for the data and task at hand~\cite{VOSSEN201660}. For other modalities, specific approaches exist. For example, for visual artifacts 
\textit{Computer Vision} is used for analysis, information extraction, as well as reconstruction and restoration. While traditionally, more feature extraction-based pipelines were used, with the recent success of (deep) neural networks for this task, such approaches are now applied to humanities datasets, leading to novel insights, but also exposing challenges around its usage~\cite{wevers2020visual}. 


\subsection{Requirements and challenges}
As argued above, since its early beginnings, DH has not only concerned itself with the application of digital methods to enhance humanities research, but also has focused on a critical reflection of data, tools, and workflows. Especially in the usage of AI tools and methods, the way that such tools collaborate with (human) scholars has been questioned. Collecting and categorizing these elements allows us to define five challenges for successful integration of AI tools in DH.  We list them below in short form and expand on the challenges and their mapping to HI capabilities in Section \ref{sec:mapping}.

\begin{description}
    \item[Embedding in scholarly practice.] AI should be embedded in the DH research methodology, through collaboration between human and artificial agents.
    \item[Data or source criticism.] Humanities data have specific charateristics that makes them different from other domain data. Moreover, a critical stance to data is a key component in the humanities and successful AI needs to support this. 
    \item[Tool criticism.] Similarly, a critical attitude towards methods and tools used is necessary. A successful AI systems supports critical inquiry into itself.  
    \item[Polyvocality and diversity.] The humanities recognizes that multiple perspectives are present and the diversity of (citizen) scientists, data and voices is key. 
    \item[Distant vs Close reading.] In the digital humanities there is ongoing debate about the relation between distant reading (shallow analysis at scale)  versus close reading (deep inquiry of text). This presents excellent opportunities for hybrid solutions. 
\end{description}

\section{Hybrid Intelligence and the CARE principles}
\label{sec:hi}
    Hybrid Intelligence as a field concerns itself with the investigation and design of human-AI ecosystems. As such, it builds on AI research, but focuses on interaction with users. \cite{akata2020research} recognizes four research challenges, abbreviated as the CARE principles. These principles can be formulated as capabilities for AI systems that are effective in HI scenarios: 
    \begin{description}
        \item[Collaborative.] AIs systems will need to be designed and constructed in such a way that they are able to cooperate in synergy with human actors in a variety of tasks. This collaboration should ultimately take into account the strengths and limitations of both the human and artificial agents. 
        \item[Adaptive] HI systems function in dynamic environments with diverse human-agent teams, necessitating adaptability to changing contexts, variable team structures, preferences, and roles. 
        \item[Responsible] Mitigating risks requires integrating ethical and legal considerations into the design and operation of  HI systems. Values like transparency, accountability, privacy, and fairness must be intrinsic to the design process and  performance.
        \item[Explainable] For mutual understanding between intelligent agents and humans, explanations are vital for shared awareness, goals, and collaborative strategies. This requires the use of methods and algorithms that are not only transparent, but can interact with users to explain their reasoning. 
    \end{description}

\section{Mapping CARE principles to DH requirements}
\label{sec:mapping}
In this section, we take the requirements and challenges as identified in Section \ref{sec:dh} and explore how the CARE principles introduced in Section \ref{sec:hi} can be applied to these challenges. Each time we present a description of the challenge;  which of the HI principles are most suited to be mapped to this case and open challenges. 

\subsection{Embedding in the Scholarly Practice}\label{sec:emb}

   In order to build good AI tools for DH that users can and will effectively adopt in their daily practice, creators of such tools need to understand the strengths and weaknesses of humans and AI alike. This requires them to \textbf{adapt} to scholars' preferences in working with and gaining understanding of historical material, and their goals, cultural biases, and social norms. 
    Common roles that scholars take in such \textbf{collaborative} efforts are that of the reader or analyst, that of the annotator in crowd or niche-sourcing~\cite{dijkshoorn2013personalized} endeavors, or that of the curator of enriched material. In their workflow, humanities scholars tend to prefer close over distant reading of source material (see Section \ref{sec:dist}), and tend to avoid systematic keyword searches. Moreover, they often work exploratory; in search of interesting pieces of information that contribute to the point they aim to make, and might need to consult sources of hundreds of years old~\cite{warwick2012digital}.
    AI systems can tailor to such needs, but creators of such tools also require understanding of how humanities scholars  perceive the output of AI systems. Errors, for instance, can create skepticism towards the capabilities of tools. For example, whereas error rates of 60-95\% for a handwriting recognition system can greatly improve searchability of archival content, Schomaker et al. mention that ``\textit{Scholars [...] are disappointed by the OCR quality of machine-printed text, which may be far above 95''}~\cite{schomaker2016design}. Moreover, humanities researchers tend to be more skeptical towards sophisticated tools, and prefer easy of use and transparency~\cite{gibbs2012building}.

   
   In an effort for natural dialogue between AI and humanities scholars, DH has started using \textbf{eXplainable} AI (XAI) techniques to generate insights or explanations of humanities datasets~\cite{el2023explainability,diaz2020accessible}. Examples may take the form of narratives, i.e.,~\cite{blin2022building,deBoer2015dive} in which data are organized according to important historical events and actors, mimicking how humans understand phenomena. Moreover, 
 interpretable rules can be used to explain associations. Wilcke et al.~\cite{wilcke2019user}, for instance, mine explainable associations from archaeological materials as novel insights for archaeologists. Such XAI tools remain limited in the humanities~\cite{el2023explainability}, but have the potential to increase interpretability of collections, as well as generate trust in human-machine collaboration.\\ 
    \textbf{HI mapping} Collaborative, Adaptive, Explainable\\
    \textbf{Open challenges} 
Which cognitive biases do scholars have in understanding humanities datasets and working with AI systems? What are natural ways of interacting with digitized material via AI systems? 

\subsection{Data or Source Criticism}\label{sec:data} 
    
        Data or source criticism deals with the critical evaluation of sources in the context of historical studies. This includes assessing the quality, completeness, bias, and representativeness of the data, as well as considering the ethical implications of data collection and use. These concern not only texts, but also other modalities such as maps, photographs and drawings~\cite{jadh13}. For humanities scholars it is obvious that historical inequalities lead to datasets that are incomplete, biased, hide certain perspectives (see Section \ref{sec:pol}) such as indigenous people's voices or certain social groups~\cite{ortolja2022encoding,van2021polyvocal,alkemade2023datasheets}, or use language that is inappropriate in today's society~\cite{brate2021capturing,nesterov2023contentious}.
        
        With use of ML, especially 
        generative AI models, 
         data criticism becomes even more important. Humanities datasets have typically not  been created as ML benchmarks, but for the purpose of knowledge communication, preservation, correspondence, or other. Moreover, they are highly heterogeneous, and their structure changes over time~\cite{alkemade2023datasheets}. 
        How can AI systems act \textbf{responsibly}, when built on top of untrustworthy sources? Other than building responsible datasets, AI systems should be capable of \textbf{explaining} the sources on which they base their reasoning in a transparent way. 

        Knowledge representation can help organize data, providing insights on knowledge gaps. Datasheets~\cite{alkemade2023datasheets}, for instance, describe datasets in a structured way, so that these can be fit more easily to the information needs of data re-users. Moreover, adhering to principles of FAIR data management \cite{wilkinson2016fair} will not only increase the reusability of digital data and methods, but also ensure that they can be subjected to the same rigorous criticism of tools and data as is common in humanities research \cite{koolen2019toward}. 
        However, when applying standard data models and reasoning for the interpretation of historical sources, there is the risk of simplification or reinterpretation of such sources in the current scientific paradigm, as socio-cultural data is inherently vague and ambiguous~\cite{jansen2019ontologies}. \\
    \textbf{HI mapping} Responsible, Explainable\\
    \textbf{Open challenges}. What is the best way to measure whether AI systems act responsibly with respect to the data that they use? Moreover, how can AI systems explain to humanities scholars the provenance of data used for explanations or insights?

\subsection{Tool Criticism}\label{sec:tool}
Related to data criticism is \textit{tool criticism}. Defined in \cite{vanEsToolCrit} as "[\textit{...] the critical inquiry of knowledge technologies considered or used for various purposes [...]}". As such, tool criticism is at the heart of the DH research methodology. It calls for critical reflection on how tools came to be, how the tool functions and the way in which digital tools are used in the digital history methodology\cite{koolen2019toward}. As an example, when using NLP tools, an uncritical with respect to the tools capabilities recognizing spelling variations or dealing with diacritical characters could lead to incorrect conclusions. 
Specifically for AI technology, the danger exists that these tools are more opaque: black-box models have a hard time explaining why certain results are produced, risking perpetuating existing biases~\cite{ortolja2022encoding} and large language models do not always allow for tracing query results back to original sources. In order to support a critical view on the use of tools in a hybrid setting, AI systems need to be transparent about their inner workings, what data they are trained on, what their limitations are and how small changes in settings influence results. This requires that notions of data provenance are integrated in such tools\cite{KusterProv,ockeloen2013biographynet}. 
The capabilities of \textbf{responsible} maps well onto this requirement as systems that collaborate with humans need to be developed in responsible ways (responsible in design). They also need to be aware of and behave in a responsible manner, specifically in being transparent. This also makes it clear why \textbf{explainable} systems are needed: When systems are able to communicate about their inner workings, they allow for the required critical inquiry by the humanities scholar.
    A special mention must be given to Generative AI, that produces novel output based on patterns learned from data. Use of this relatively new form of AI is starting to be explored in the DH context. Examples include using Generative Adversarial Networks to analyse visual styles \cite{offert2020generative} or generate images of medieval manuscripts depicting steam trains\cite{makingfaking}.  Retrieval Augmented Generation (RAG) combines an information retrieval component with text generation based on large language models\cite{garcia2023if}. When responses or 'alternative' texts, images or other sources can be generated convincingly, it becomes even more important that such methods are transparent.
    \\
    \textbf{HI mapping.} Responsible, Explainable\\
    \textbf{Open challenges.} One of the main challenges here is how state-of-the-art ML can be made 'criticisable'. Here the field of explainable AI (XAI) can provide opportunities, but for successful integration, more research is needed into how DH scholars can interact with systems in order to retrieve suitable explanations. At the same time, standards for describing and sharing tool descriptions are needed, building on \cite{koolen2019toward}.
    
\subsection{Polyvocality and Diversity}\label{sec:pol}
    In the humanities, the concept of \textit{positionality} is often used to indicate that the research process itself as well as the researcher are not neutral, but bring their own perspective and biases\cite{bourke2014positionality}. \textit{Polyvocality} is essential in digital humanities as it promotes a diverse and inclusive approach to the study of cultural artifacts and human experiences in the digital realm\cite{schofield2018co,van2021polyvocal}. By incorporating a variety of voices and perspectives, polyvocality helps mitigate biases, avoid stereotypes, and reflect the complexity of the real world. It enhances interdisciplinary collaboration, engages communities, and contributes to more comprehensive representations of cultural heritage\cite{golding2013museums}. Where platforms facilitate public engagement, polyvocality becomes crucial for fostering a more inclusive dialogue about culture, history, and society. Overall, embracing multiple voices in digital humanities enriches research, analysis, and understanding, leading to a more nuanced exploration of the complexities inherent in human narratives and cultural phenomena.
    This can refer not only to \textit{diversity} in the collection, but also of the persons involved in the research activities, including experts and lay persons. This need for polyvocality connects well with the goals of \textbf{responsible} AI, specifically where it concerns fairness towards various representations and the way in which systems deal with bias. In DH practice, this has for example the implication that methods used for annotating datasets (be they automatic, through crowdsourcing or hybrid solutions) will need to be able to capture, represent and present such diverse perspectives\cite{van2021polyvocal}. This requires that user-facing methods are aware of and can deal with diverse end-users. This diversity can be in the expertise dimension, but also in the cultural or age dimension. \textbf{Collaborative} AI is needed, specifically AI that is able to collaborate with a variety of users. In that sense it will have to be \textbf{adaptive} to be able to deal with multiple perspectives, various cultural contexts and be able to keep 'disagreement' between experts, laypersons and algorithms intact.\\
    \textbf{HI mapping}  Collaborative, Adaptive, Responsible\\
    \textbf{Open challenges} Diversity of users will be a key challenge for successful HI tools that support or even promote polyvocality. Many AI systems and interfaces are designed for specific end-users, be they domain experts or laypersons. They often are accessible only to a limited number of people, and might require considerable resources. Systems that cater to diverse users 
    will be needed to make sure all relevant voices can be heard. 

\subsection{Distant and Close Reading}\label{sec:dist}
    Taken from literary studies, \textit{Distant reading} and \textit{close reading} represent two contrasting approaches to humanities research. Close reading involves meticulous examination and interpretation of individual texts, focusing on details like language, imagery, and structure to uncover deeper meanings and themes. Conversely, distant reading employs methods to analyze large corpora of texts, aiming to identify broader patterns, trends, and relationships across works. It emphasizes more quantitative analysis, data visualization, and the exploration of phenomena at scale\cite{moretti2013distant}. While traditional humanities research often includes much close reading, with the rise of computational methods, in DH, more interest for distant reading has emerged for a variety of modalities\cite{janicke2015close}. Recognizing this complementarity paves the way to use \textbf{collaborative} AI to the fullest effect: HI systems will need to collaborate with human scholars and be flexible enough to switch between supporting distant reading, but also supporting close reading. They will need to be \textbf{adaptive} in the sense that the results of close reading will be needed to be taken into account for further processing.  
    Pattern recognition using Machine Learning can significantly enhance this distant reading and connecting large numbers of sources through knowledge graphs allows for querying and analysis of large interconnected corpora. Human experts can then engage more deeply with smaller amounts of objects of study. For example, the CLARIAH media suite specifically is designed to support both this distant reading (using AI to connect audiovisual content) and close reading (providing easy access and annotation tools for individual videos)\cite{martinez2017tools}. 
    Where now, these activities are disparate, \cite{bonfiglioli} recognize that further exploitation of the  complementarity between computational methods and humans is a worthwile research direction. Such deeper collaboration between humans and machines fits the research agenda and capabilities of HI well. \\
    \textbf{HI mapping.} Collaborative, Adaptive  \\
    \textbf{Open challenges.} To support the seamless transition between distant and close reading, AI systems need to maintain connections between aggregations and the original sources. LLMs for example, often only store the statistical aggregations and cannot trace back query answers to specific texts. Knowledge Graphs that connect various data sources now often only include metadata, and not the actual documents, images or videos. Integral solutions are needed that actually contain such multimodal data (e.g. \cite{wilcke2017knowledge}).

\section{Discussion}
In the preceding section, we have outlined how the capabilities of HI match with the requirements of Digital Humanities. HI can make deep collaborations between AI systems and with human scholars possible. It can be truly successful if such systems are designed taking into account the diversity of the users and methodological concerns that they have concerning tools and data. This calls for systems that deal well with provenance and transparency and are responsible by design. 

While we argue here that DH as a field can benefit from the Hybrid Intelligence initiative, we also recognize that the humanities provide HI not just as a playing ground for interesting and difficult challenges regarding human-AI collaboration, but also that for fundamental AI/HI challenges, the field can learn much from the humanities. Questions such as how to deal with multiple perspectives, how to conceive of and deal with social bias and (algorithmic) fairness should not be answered solely by AI experts, but can be addressed more deeply by taking into account the decades of experience that exists in the humanities addressing such issues. With the increased impact of AI in society and the volume of current public debate about its ethics, AI in general, and HI in specific, is in need of humanists ``\textit{are ideal “domain experts” for the current juncture}'' \cite{goodlad}.

As Hybrid Intelligence is a current research initiative, many of the mappings identified above are future research directions. However, we here end our paper with two specific overarching challenges towards HI for DH. First of all, a danger in Digital Humanities is that of one-shot solutions. Tools and methods are often designed for a single (research) project, and lack in reusability across collections or subfields. There should be a focus on reusable, Open Source, tools, datasets and methods that are well described (see Tool and Data criticism) and usable for a variety of end-users. This usability also requires careful development of user-interfaces that are well-integrated in the humanities task at hand. This can include web-portals,, dialogue systems or interactive notebooks~\cite{viola}.

Secondly, although we here have been mostly focusing on how HI systems should be designed, of course the human scholars will need to play their part. In order to have truly effective Hybrid teams, the human actors will also need to be able to be critical and cooperative with their computational team mates. This requires not only an open mind, but an increased level of AI literacy. Including digital methods in the humanities curricula will be crucial to ensure that the next generation of humanities scholars is ready to play their part in the Hybrid Intelligent team.
\\
\\
\noindent
\textbf{Acknowledgements.} This research was supported by the Hybrid Intelligence Center, funded  through the Netherlands Organisation for Scientific Research (NWO), \url{https://hybrid-intelligence-centre.nl} grant number 024.004.022 and by the Pressing Matter project (\url{http://pressingmatter.nl}), funded by NWO through the Netherlands National Research Agenda (NWA.1292.19.419).




\bibliographystyle{splncs04}
\bibliography{bibliography}

\end{document}